\documentclass[preprint,aps,showpacs,nofootinbib,preprintnumbers,amsmath,amssymb]{revtex4-1}
\usepackage{epsfig}
\usepackage{bm}% bold math
\usepackage[usenames ,dvipsnames]{xcolor}
\usepackage{slashed}
\usepackage{graphicx,color}

\usepackage{multirow}
\begin{document}
	\title{Mixing  effects  of $\Sigma^0-\Lambda^0$ in $\Lambda_c^+$ decays}
		
\author{C.Q. Geng$^{1,2,3,4}$, Chia-Wei Liu$^{3}$ and Tien-Hsueh Tsai$^{3}$}
\affiliation{
$^{1}$School of Fundamental Physics and Mathematical Sciences, Hangzhou Institute for Advanced Study, UCAS, Hangzhou 310024, China \\
$^{2}$International Centre for Theoretical Physics Asia-Pacific, Beijing/Hangzhou, China \\
$^{3}$Department of Physics, National Tsing Hua University, Hsinchu 300, Taiwan\\
$^{4}$Physics Division, National Center for Theoretical Sciences, Hsinchu 300, Taiwan
}\date{\today}

	\begin{abstract}
We analyze the mixing between $\Sigma^0$ and $\Lambda^0$ based on the baryon masses.
We distinguish the contributions from QCD and QED in the baryon mass splittings.
We find that the mixing angle between  $\Sigma^0$  and $\Lambda^0$  is  $(2.07\pm 0.03)\times 10^{-2} $, 
which leads to  the decay  branching fraction and up-down asymmetry 
of  $\Lambda_c^+ \to \Sigma^0 e^+ \nu_e$ 
 to be  ${\cal B}(\Lambda_c^+ \to \Sigma^0 e^+ \nu_e)=(1.5\pm 0.2)\times 10^{-5}$ and $\alpha(\Lambda_c^+ \to \Sigma^0 e^+ \nu_e)=-0.86\pm 0.04$, respectively.
Moreover, we obtain that  $\Delta {\cal B}\equiv {\cal B}(\Lambda_c^+\to \Sigma^0 \pi^+) - {\cal B}(\Lambda_c^+\to \Sigma^+\pi^0)=(3.8\pm 0.5)\times 10^{-4}$ and $\Delta \alpha \equiv\alpha(\Lambda_c^+\to \Sigma^0 \pi^+) -\alpha(\Lambda_c^+\to \Sigma^+\pi^0)=(-1.6\pm 0.7)\times10^{-2}$, which should vanish without the mixing.
	\end{abstract}

\maketitle
\section{INTRODUCTION}
The flavor content is one of the cornerstones in particle physics. The hadrons are often categorized and named in terms of their flavor states. For instance, the hadrons in the isospin group, such as pions,  share the same name. Since the QCD energy scale is much larger than the mass difference of the hadrons with the same isospin, it is believed that
the hadrons also have similar wave functions. The isospin group can be extended to $SU(3)$ by including the strange quark,
which is the so-called $SU(3)_F$ flavor symmetry and has been widely used in particle physics.

Based on the $SU(3)_F$ symmetry, the precision for the  Gell-Mann-Okubo (GMO) mass formula is around one percent, indicating that the baryons indeed share the similar wave functions.
Another well know mass formula based on $SU(3)_F$ for the octet baryons is the Coleman-Glashow~(CG) one~\cite{Coleman:1961jn}. Note that the masses of the octet baryons have been intensively studied  in the calculations of the Lattice QCD~(LQCD)~\cite{latticeQCD1,latticeQCD2,latticeQCDQED1,latticeQCDQED2,latticeQCDQED3} as well as the theoretical models with the baryon wave functions~\cite{Isgur:1979ed,theo1,theo2,theo3,theo4,theo5,theo6CG}. 

The octet baryon states of $\Sigma^0$ and $\Lambda^0$ have the same quark components of $uds$. 
Originally, they are categorized by the isospin property with  $\Sigma^0$ and $\Lambda^0$ being the   triplet and singlet states
under the $SU(2)_I$ isospin symmetry, respectively. This categorization is based on  
that the isospin symmetry is much better than  $SU(3)_F$. However, 
both $SU(2)_I$  and $SU(3)_F$ are not exact, resulting in a possible mixing between $\Sigma^0$ and $\Lambda^0$. 
The physical baryons shall be made of the mixing of isospin triplet and singlet states. In general, the mixing angle is estimated 
to be the ratio of the  $SU(2)_I$
 and  $SU(3)_F$ breaking energy scales.  Note that the calculation in the LQCD gives the mixing angle $\theta = 0.006 \pm 0.003$~\cite{Horsley:2014koa}.
 
 %\textcolor{blue}{
 Recently, the BES\MakeUppercase{\romannumeral 3} Collaboration has announced the up-down asymmetries for $\Lambda_c^+ \to \Sigma \pi$, given by~\cite{Ablikim:2019zwe}
 \begin{eqnarray}
 \label{q1}
 \alpha(\Lambda_c^+ \to \Sigma^+ \pi^0)&=&-0.73 \pm 0.18 \,,\nonumber\\
 \alpha(\Lambda_c^+ \to \Sigma^0 \pi^+)&=&-0.57\pm 0.12\,.
 \end{eqnarray}
 Note that 
  the corresponding branching ratios have been measured to be~\cite{pdg}:
  % given by
 \begin{eqnarray}
 \label{q2}
 {\cal B}(\Lambda_c^+ \to \Sigma^+ \pi^0)&=&(1.24\pm 0.10) \%\,,\nonumber\\
{\cal B}(\Lambda_c^+ \to \Sigma^0 \pi^+)&=&(1.29\pm 0.07) \%\,.
 \end{eqnarray}
 However, in the limit of the isospin symmetry, both asymmetries in Eq.~(\ref{q1}) and branching ratios in Eq.~(\ref{q2})
 should be equal.
%They shall be equal in the limit of isospin symmetry. 
On the other hand, the semi-leptonic decay of  $\Lambda_c^+ \to \Sigma^0 l^+ \nu_l$ is forbidden since $\Lambda_c^+$ and $\Sigma^0$ belong to different isospin representations.
  In this study, we explore the isospin breaking effect in the $\Sigma^0-\Lambda^0$ mixing and discuss the  effects in $\Lambda_c^+$ decays.
%  }

Our paper is organized as follows. In Sec.~\MakeUppercase{\romannumeral 2}, we introduce the hadron representations under the $SU(3)_F$ flavor group. The mixing effects in $\Lambda_c^+$ decays are studied in Sec.~\MakeUppercase{\romannumeral 3}. We present our conclusions in Sec.~\MakeUppercase{\romannumeral 4}.

\section{Hadron Representations in SU(3)$_{\text {F}}$}
In terms of $SU(3)_F$,  the matrix representations of the octet baryons can be written as~\cite{georgi}
\begin{eqnarray}\label{OctetBaryon}
({\bf B}_n)^i_j&=&\left(\begin{array}{ccc}
\frac{1}{\sqrt{6}}\Lambda^{\prime 0}+\frac{1}{\sqrt{2}}\Sigma^{\prime 0} & \Sigma^+ & p\\
\Sigma^- &\frac{1}{\sqrt{6}}\Lambda^{\prime 0} -\frac{1}{\sqrt{2}}\Sigma^{\prime 0}  & n\\
\Xi^- & \Xi^0 &-\sqrt{\frac{2}{3}}\Lambda^{\prime 0}
\end{array}\right)_{ij}\,,
\end{eqnarray}
where the prime “$\prime$”  denotes the un-mixed state.
For instance, the proton's matrix representation and state correspond to $(p)^i_j=\delta^{i1}\delta_{j3}$ and  $(p)^i_j|^j_i\rangle=|^3_1\rangle$, respectively.

In the standard model, the $SU(3)_F$ symmetry is broken by the quark masses as well as  the electromagnetic interaction. The matrix representations of the light quark masses and electric charges of the quark flavors are given as
\begin{eqnarray}\label{Matrix}
M=\left( \begin{array}{ccc}
m_u&0&0\\
0&m_d&0\\
0&0&m_s\\
\end{array}\right)-\frac{1}{3}(m_u+m_d+m_s)
\,\,\,\,\text{and}\,\,\,
Q=\frac{1}{3}\left( \begin{array}{ccc}
2&0&0\\
0&-1&0\\
0&0&-1\\
\end{array}\right)\,,
\end{eqnarray}
respectively. Note that both $M$ and $Q$ belong to ${\bf 8}$ under the $SU(3)_F$ group.
Consequently, the baryon mass must be the function of $M$ and $Q$, given by
\begin{equation}\label{MassElement}
M_B= \langle B | H(M,Q)| B \rangle \,,
\end{equation}
where $H$ and $|B\rangle$ are the mass operator and state of the octet baryon, respectively.

Naively, one may write down the baryon mass operator as
\begin{equation}\label{Naive}
H(M,Q)=H^0+	H^1_m(M) + H^1_q(Q) + O(H^2)\,,
\end{equation}
where  the superscripts of $n=0,1,2$ stand for the $n$-order approximations and the subscripts of $m$ and $q$ imply the breaking sources of the $SU(3)_F$ symmetry.  However, the second-order correction from the strange quark mass can be the same size as the first-order one from the up and down quark masses, e.g. $(m_s/\mu_H)^2 \sim m_{q}/\mu_H$, where $\mu_H$ is the typical hadronic  scale.

A better way to do the approximation is to categorize the breaking effects according to their symmetry properties instead of the sources. We rewrite Eq.~\eqref{Matrix} as
\begin{eqnarray}\label{decomposition}
&&M=  m_{\overline{s}}T_8+  m_{\overline{q}} T_3 \,\,\,\,\,\text{and}\,\,\,\,\,Q=\frac{1}{6}T_8 + \frac{1}{2}T_3\,, 
\end{eqnarray}
with
\begin{eqnarray}
&&
T_8=\left( \begin{array}{ccc}
1&0&0\\
0&1&0\\
0&0&-2\\
\end{array}\right)
\,,\,\,\,\,\,\,
T_3=\left( \begin{array}{ccc}
1&0&0\\
0&-1&0\\
0&0&0\\
\end{array}\right)\,, 
\end{eqnarray}
where $ m_{\overline{s}}=-\left( 2 m_s-m_u-m_d\right)/6$ and $ m_{\overline{q}}=\left( m_u - m_d\right)/2 $. Here, we have decomposed the matrix representation into two different parts. Note that $T_8$ is invariant under the isospin transformation, whereas $T_3$ is not.
Accordingly, the baryon mass operator in Eq.\eqref{Naive} is given by 
\begin{eqnarray}\label{MassWeUse}
&&H(M,Q)=H^0+	H^1_m\left( m_{\overline{s}}T_8+  m_{\overline{q}} T_3 \right) + H^1_q\left(\frac{1}{6}T_8 + \frac{1}{2}T_3\right) + O(H^2)\nonumber\\
&&= H^0+	\left(H^1_m(  m_{\overline{s}}T_8) + H^1_q(\frac{1}{6}T_8)\right) + \left(H^1_m( m_{\overline{q}}T_3) +H^1_q( \frac{1}{2}T_3) \right) + O(H^2)\nonumber\\
&&=H^0 + H^1_S(T_8) + H^1_I(T_3)+O(H^2)\,,
\end{eqnarray}
where we have used that the matrix representation is linear and the operators in the parentheses have the same representation, respectively.
Since $H^1_S$ contains the correction from $m_s$, it is much larger than $H^1_I$. 
On the other hand, the second-order correction in $O(H^2)$ caused by $m_s$ can be the same order as $H^1_I$. Explicitly, we have the hierarchy, given by 
\begin{equation}
H^0 \gg H^1_S \gg H^1_I \sim O(H^2)\,.
\end{equation}
Note that the correction from  $m_s$  is invariant under the isospin transformation.
If $O(H^2)$ is neglected in the calculation of $H^1_I$, it is only reasonable to deal with the physical quantities, which are not affected by the correction from $m_s$. 

Notice that the baryon wave functions in Eq.~\eqref{OctetBaryon} are chosen as the eigenvectors of $H^0+H^1_S$. Explicitly, they have the following properties
\begin{equation}\label{HSdiagonal}
\langle B^\prime| H^0+H^1_S|B \rangle = \langle B^\prime| B \rangle =0  \,\,\,\,\,\,\,\,\,\, \text{if} \,\,\,\,\,\,B^\prime \neq B\,.
\end{equation}
However, the physical baryon states are the eigenvectors of the full baryon mass operator instead.
 In general, the isospin breaking term $H^1_I$ has a nonzero matrix element between $\Sigma^{ \prime 0 }$ and $\Lambda^{ \prime 0}$, i.e. $\langle \Sigma^{ \prime 0}| H^1_I| \Lambda^{\prime 0}\rangle\neq 0$, while the correction due to $m_s$  has no contribution to this matrix element due to the isospin symmetry.
The physical baryon condition is given by
\begin{eqnarray}\label{Full}
&&\langle B_P^\prime| H |  B_P \rangle=0 \,\,\,\,\,\,\,\,\,\, \text{if} \,\,\,\,\,\, B_P^\prime \neq B_P\,, 
\end{eqnarray}
with~\cite{Mixing_El,Mixing_Eq}
\begin{eqnarray}\label{Full2}
&&|\Lambda^0\rangle=\cos\theta |\Lambda^{0 \prime} \rangle- \sin \theta |\Sigma^{0 \prime}\rangle\,\,\,\,\, \text{and}\,\,\,\,\, |\Sigma^{ 0}\rangle=\cos \theta |\Sigma^{0 \prime }\rangle + \sin \theta |\Lambda^{0 \prime}\rangle\,,
\end{eqnarray}
where $P$  denotes as the physical baryon  and $\theta$ is the mixing angle to be determined through the octet baryon masses. 

The baryon masses are evaluated from Eqs.~\eqref{MassElement} and \eqref{MassWeUse}. With the $SU(3)_F$ symmetry described in the beginning of the previous section, we can apply the Wigner–Eckart theorem. Consequently, the matrix element in Eq.~\eqref{MassElement} is parametrized as
\begin{eqnarray}\label{firstEq}
\langle B^\prime|H^0+H^1_S|B\rangle&=&M_0+S_1\text{Tr}~(\overline{{\bf B}^\prime}_nT_8{\bf B}_n) +S_2\text{Tr}~(\overline{{\bf B}^\prime}_n{\bf B}_nT_8)\,,\\
\label{secondEq}
\langle B^\prime| H^1_I+O(H^2)|B \rangle&=&I_1\text{Tr}~(\overline{{\bf B}^\prime}_nT_3{\bf B}_n) +I_2\text{Tr}~(\overline{{\bf B}^\prime}_n{\bf B}_nT_3)+O(H^2)\,,
\end{eqnarray}
where $M_0$, $S_i$ and $I_i$ are the $SU(3)_F$ parameters to be extracted through the experiments, $O(H^2 )$ is the higher order correction, and $\overline{{\bf B}}_n\equiv{\bf B}_n^\dagger$. In fact, Eq.~\eqref{firstEq} corresponds to the GMO mass formula, while Eq.~\eqref{secondEq} can give us the CG one.
% In this work, we would derive the CG mass formula in a more systematic way together with the GMO one.
To get a consistent result, the parameters of $I_1$ and $I_2$ are determined with the mass differences between the isospin related baryons,  
while $M_0$, $S_1$ and $S_2$ are fixed with the average baryon masses within the same isospin subgroup.

 There is one  triplet ($\Sigma$), two  doublets ($\Xi$ and $N$) and one  singlet ($\Lambda^{\prime 0  }$)  under $SU(2)_I$
 in the octet baryons. Their experimental mass differences and $SU(3)_F$ parametrizations are summarized in Table \ref{Table1}, where the results from QCD and QED are given at the end of this section. We have rounded the experimental data to the second decimal place due to that the $SU(3)_F$ symmetry is only an approximation. The higher precision is not expected.
\begin{table}[h]
	\caption{Mass differences~(MeV) in  the octet baryons and $SU(3)_F$ parametrizations with the experimental data and our results from QCD and QED.
}	
	\begin{center}
\begin{tabular}[t]{c|cccc}%\label{Table12}
	\hline
	Mass differences&$SU(3)_F$ parameters&data~\cite{pdg}&QCD & QED\\
	\hline
$M_{\Sigma^-}-M_{\Sigma^+}$&$2I_1-2I_2$&$8.08\pm 0.08$&$8.59\pm 0.01$&$-0.51\pm 0.08$\\
$M_{\Sigma^-}-M_{\Sigma^0}$&$I_1-I_2$&$4.81\pm 0.04$\footnote{We have used the approximation of  $M_{\Sigma^0}=M_{\Sigma^{\prime 0}}$}&$4.30\pm 0.00$&$-0.26\pm 0.04$\\
$M_n-M_p$&$-2I_2$&$1.29\pm 0.00$ &$2.84\pm 0.00$&$-1.54\pm 0.00$\\
$M_{\Xi^-}-M_{\Xi^0}$&$2I_1$&$6.85\pm 0.21$&$5.76\pm 0.00$&$1.04\pm 0.08$\\
\hline
\end{tabular}
\end{center}
\label{Table1}
 \end{table}

From Table~\ref{Table1}, one can easily obtain the CG mass formula, given by
\begin{equation}\label{CG}
(M_{\Sigma^-}-M_{\Sigma^+})-(M_n-M_p)=M_{\Xi^-}-M_{\Xi^0}\,,
\end{equation}
along with additional one, given by
\begin{equation}\label{add}
	\frac{1}{2}(M_{\Sigma^-}-M_{\Sigma^+})=M_{\Sigma^-}-M_{\Sigma^{\prime 0}}\,.
\end{equation}
If one assumes that the baryon mass operator is invariant under  $SU(2)_I$, 
the CG mass formula is trivial since the baryons within the same isospin subgroup should have the same mass. Consequently, the CG mass formula is often interpreted as the equation between the isospin mass differences.
 In addition, it can be related to the correction in the U-spin symmetry of the down and strange quarks as well due to the rearrangement of
\begin{equation}\label{trivial}
(M_{\Sigma^-}-M_{\Xi^-})+(M_p-M_{\Sigma^+})+(M_{\Xi^0}-M_{n})=0\,,
\end{equation}
where $(\Sigma^- , \Xi^-)$, $(p, \Sigma^+)$ and $(\Xi^0, n)$ belong to the same U-spin subgroup.
In fact, Eq.~\eqref{trivial} is trivial in the original derivation in Ref.~\cite{Coleman:1961jn} based on  the mass operator being only a function of Q, which is invariant under the U-spin transformation, e.g. $M_{\Sigma^-}=M_{\Xi^-}$, $M_p=M_{\Sigma^+}$ and $M_{\Xi^0}=M_{n}$. 
Moreover,  the CG mass formula can also be related to the correction 
under  the V-spin symmetry of the up and strange quarks with the rearrangement of
\begin{equation}\label{trivialU}
(M_{\Sigma^-}-M_{n})+(M_p-M_{\Xi^-})+(M_{\Xi^0}-M_{\Sigma^+})=0\,,
\end{equation}
where $(\Sigma^- , n)$, $(p, \Xi^-)$ and $(\Xi^0, \Sigma^+)$ have the same V-spin representations.
Likewise, Eq.~\eqref{trivialU} is trivial if the V-spin symmetry is exact in the baryon mass operator.
To sum up, the CG mass formula is automatically satisfied if one of the $SU(2)$ subgroups in $SU(3)_F$ is preserved.

From the experimental data of $M_{\Sigma^-}-M_{\Sigma^+}$ and $M_n-M_p$, the parameters $I_1$ and $I_2$ in Eq.~\eqref{secondEq} can be found to be
\begin{equation}\label{ISOSPIN}
I_1=\left(3.40\pm 0.04\right) \text{MeV}\,\,\,\,\,\,\text{and}\,\,\,\,\,\,I_2=\left(-0.65\pm 0.00\right) \text{MeV} \,.
\end{equation}

Let us now focus on $H^1_S$. To get consistent results cooperated with $H^1_I$, we  use the average masses in the isospin subgroups, defined as $M_\Sigma=(M_{\Sigma^+}+M_{\Sigma^-})/2$, $M_N=(M_p+M_n)/2$, and $M_\Xi=(M_{\Xi^-}+M_{\Xi^0})/2$. The famous GMO mass formula can be  obtained through Eq.~\eqref{firstEq}, read as
\begin{equation}
\label{q21}
M_{ \Lambda^0 } \approx M_{\Lambda^{\prime 0}}={1\over 3}\left[2(M_\Xi + M_N)-M_\Sigma   \right]\,.
\end{equation}
%In the first equation, we have approximated $M_{\Lambda_P^0}\approx M_{\Lambda^0}$.
Experimentally, the left and right hand sides of Eq.~(\ref{q21})
are given by $1115.68\pm 0.01$ and $1107.00\pm 0.07$~MeV, respectively.  In addition,  we get
\begin{equation}\label{STRANGESPIN}
M_0=(1150.21\pm 0.04)~\text{MeV}\,,\,\,\,\,\,S_1=(84.83\pm 0.01)~\text{MeV}\,,\,\,\,\,\,S_2=(-41.63\pm 0.04)~\text{Mev}\,
\end{equation}
in Eq.~\eqref{firstEq}, where the experimental data of $M_\Sigma$, $M_N$, and $M_\Xi$ have been used.

If the QED effect is ignored, the baryon mass depends on the quarks mass matrix only, resulting in that
\begin{eqnarray}\label{QCD}
M_{\text{QCD}}&=& M_0+M_1\text{Tr}~(\overline{{\bf B}}^\prime_nM{\bf B}_n) +M_2\text{Tr}~(\overline{{\bf B}}^\prime_n{\bf B}_n M)+O(H^2)\,,
\end{eqnarray}
where $M_1$ and $M_2$ are the unknown parameters. By using Eqs.~\eqref{decomposition}, \eqref{firstEq} and  \eqref{secondEq}, we obtain that
\begin{equation}\label{firstRatio}
\left(\frac{I_1}{I_2}\right)_{\text{QCD}}=\left(\frac{S_1}{S_2}\right)_{\text{QCD}}\,,
\end{equation}
where the subscript of ``QCD'' indicates that the QED effect is ignored. Since $S_1$ and $S_2$ are dominated by the strange quark mass correction, we can safely approximate Eq.~\eqref{firstRatio} as 
\begin{equation}\label{firstRatio2}
\left(\frac{M_{\Xi^-}-M_{\Xi^0}}{M_n-M_p}\right)_{\text{QCD}}=-\left(\frac{I_1}{I_2}\right)_{\text{QCD}}=-\left(\frac{S_1}{S_2}\right)_{\text{QCD}}\approx-\frac{S_1}{S_2}=2.038\pm 0.002\,.
\end{equation}
The calculation in the LQCD is indeed consistent with Eq.~\eqref{firstRatio2}, where the ratio is around $1.6\sim 2.2$~\cite{latticeQCD1,latticeQCD2,latticeQCDQED1,latticeQCDQED2,latticeQCDQED3}.
In reality, $I_1/I_2$ has the value  of $-5.25\pm 0.06 $, which clearly indicates that $I_1/I_2\neq (I_1/I_2)_{\text{QCD}}$. 
Moreover, from Eq.~\eqref{QCD}, we have
\begin{equation}\label{QCDratio}
\left(\frac{I_1}{S_1}\right)_{\text{QCD}}=\left(\frac{I_2}{S_2}\right)_{\text{QCD}}=\frac{m_{\overline{q}}}{m_{\overline{s}}}=0.034\,,
\end{equation}
where  the mass ratios among the light quarks in Ref.~\cite{Weinberg:1977hb} have been used. With Eqs.~\eqref{ISOSPIN},  \eqref{STRANGESPIN} and \eqref{QCDratio} along with $I_i=(I_i)_{\text{QCD}}+(I_i)_{\text{QED}}$, we can separate the contributions to the mass differences of the isospin breakings in QCD and QED as listed in Table~\MakeUppercase{\romannumeral 1}. The results are fairly closed to those in the literature based on LQCD and QED~\cite{latticeQCDQED1,latticeQCDQED2,latticeQCDQED3}.

\section{$\Lambda_c^+$ deacys with $\Sigma^0-\Lambda^0$ Mixing}
%\textcolor{blue}{
The mixing angle is determined by~\cite{Mixing_Eq}
\begin{equation}
\tan \theta=\langle \Sigma^{ \prime 0}| H^1_I| \Lambda^{ \prime 0}\rangle /\left(M_{\Sigma^0}-M_{\Lambda^0}\right)
\end{equation}
where the matrix element is given as 
\begin{equation}
\langle \Sigma^{ \prime 0}| H^1_I| \Lambda^{ \prime 0}\rangle = \frac{1}{\sqrt{3}}\left( I_1 + I_2 \right) = 1.59\pm 0.02~ \text{MeV}\,.
\end{equation}
By using $I_1 +I_2 =\left(M_{\Sigma^{0}}-M_{\Sigma^{+}}\right)-\left(M_{n}-M_{p}\right)$, our formula is the same as the one in Ref.~\cite{Mixing_El},
in which only  the mixing through the electromagnetic interaction is considered.
%}
Consequently,
we obtain the mixing angle
\begin{equation}\label{Mixingangle}
\theta=(2.07 \pm 0.03)\times 10^{-2}\,.
\end{equation}

Although the mixing has a little effect on the baryon masses, it plays a significant role in the decays associated with the isospin symmetry. We  now study the mixing effects in the semileptonic 
and nonleptonic charmed baryon decays.
In general, the states of $\Lambda_c^+$ and $\Sigma_c^+$, corresponding to $I=0$ and $I=1$, are also mixed. However, the mixing is suppressed by the charmed quark mass~\cite{Franklin:1981rc}. As a result, it will be ignored   in this work.

Since $\Lambda_c^+$~($\Sigma^{\prime 0}$ ) is anti-symmetric~(symmetric) in up  and down quarks,
the decay channel of $\Lambda_c^+ \to \Sigma^{  0} e^+ \nu_e$ without the mixing is forbidden. The ratio between the decay rates of 
 $\Lambda_c^+ \to \Lambda^0 e^+ \nu_e$  and  $\Lambda_c^+ \to \Sigma^0 e^+ \nu_e$  is given by
 \begin{equation}
 \frac{\Gamma(\Lambda_c^+ \to \Sigma^0 e^+ \nu_e)}{\Gamma(\Lambda_c^+ \to \Lambda^0 e^+ \nu_e)}=\tan^2 \theta=(4.3 \pm 0.1)\times 10 ^{-4}\,,
 \end{equation}
 where we have approximated that $M_{\Sigma^0}\approx M_{\Lambda^0}$ in the kinematic phase space. 
 In addition, the angular distributions of the $\Sigma^0$ and $\Lambda^0$ modes should be the same.  With the experimental data for $\Lambda_c^+\to \Lambda^0 e^+ \nu_e$~\cite{pdg} and the mixing angle in Eq.~\eqref{Mixingangle}, the branching ratio and up-down asymmetry  of $\Lambda_c^+ \to \Sigma^0 e^+ \nu_e$ are given by 
 \begin{equation}\label{semilep}
{\cal B}(\Lambda_c^+ \to \Sigma^0 e^+ \nu_e)=(1.5\pm 0.2) \times 10^{-5}\,\,\,\,\,\text{and}\,\,\,\,\,\alpha(\Lambda_c^+ \to \Sigma^0 e^+ \nu_e)=-0.86\pm 0.04\,,
 \end{equation}
 respectively.
 
 We now explore the non-leptonic charmed baryon decays of $\Lambda_c^+ \to \Sigma^+ \pi^0$ and $\Lambda_c^+ \to \Sigma^0 \pi^+$. If there is no mixing between $\Lambda^0$ and $\Sigma^0$, two decays should have the same decay width and up-down asymmetry parameter, given by~\cite{Geng:2019xbo}
 \begin{eqnarray}
  \Gamma &=&\frac{p_{\Sigma}}{8\pi}\left(\frac{(m_{\Lambda_c^+}+m_{\Sigma})^2-m_\pi^2}{m_{\Lambda^+_c}^2}|A|^2
 +\frac{(m_{\Lambda^+_c}-m_{\Sigma})^2-m_\pi^2}{m_{\Lambda_c^+}^2}|B|^2\right)\,\,,\nonumber\\
    \alpha &=& \frac{2\kappa \text{ Re}(A^*B)}{|A|^2+\kappa^2|B|^2}\,,\;\;
 % \nonumber\\
 \kappa = \frac{p_{\Sigma}}{E_{\Sigma}+m_{\Sigma}}
 \end{eqnarray}
respectively, where $A$ and $B$ are associated with the P and S wave amplitudes, and $m_i$, $p_i$ and $E_i$ are the mass, momentum and energy for the $i$th hadron in the CM frame, respectively. 
On the other hand, the P wave amplitude for $\Lambda_c^+ \to \Sigma^0\pi^+$ can be written as
 \begin{eqnarray}\label{f1}
A(\Lambda_c^+ \to \Sigma^0\pi^+)&=&\cos \theta A(\Lambda_c^+ \to \Sigma^{\prime 0}\pi^+)+\sin \theta A(\Lambda_c^+ \to \Lambda^{\prime 0}\pi^+)\nonumber\\
&=& -\cos \theta A(\Lambda_c^+ \to \Sigma^+\pi^0)+\sin \theta A(\Lambda_c^+ \to \Lambda^{\prime 0}\pi^+)\,,
 \end{eqnarray}
 where  the second equation is given by the isospin symmetry~\cite{Geng:2019xbo}. 
 The S wave amplitude can be given by
 replacing $A$ with $B$ in Eq.~(\ref{f1}). 
 From the experimental values of ${\cal B}(\Lambda_c^+ \to \Sigma^+ \pi^0)=(1.24\pm 0.10)\%   $ and $\alpha(\Lambda_c^+ \to \Sigma^+ \pi^0)=-0.73\pm 0.18$~\cite{pdg,Ablikim:2019zwe}, we obtain that\footnote{There are four different solutions for $A$ and $B$ with given values of ${\cal B}$ and $\alpha$. We have chosen the solution that is consistent with the $SU(3)_F$ analysis~\cite{Geng:2019xbo}.}
 \begin{eqnarray}\label{f2}
 &&A(\Lambda_c^+ \to \Sigma^+ \pi^0)=(5.7\pm 0.4)10^{-2} G_F\text{GeV}^2 \,, \nonumber\\
 &&B(\Lambda_c^+ \to \Sigma^+ \pi^0)=(-7.9\pm 2.4)10^{-2} G_F\text{GeV}^2\,,
 \end{eqnarray}
with the correlation of $R=0.73$.
Similarly, with the data of ${\cal B}(\Lambda_c^+ \to \Lambda^0 \pi^+)=(1.30\pm 0.07)\%$ and $\alpha(\Lambda_c^+ \to \Lambda^0 \pi^+)=-0.80\pm 0.11$, we find that$^1$
 \begin{eqnarray}\label{f3}
 &&A(\Lambda_c^+ \to \Lambda^{\prime 0} \pi^+)=(-2.9\pm 0.5)10^{-2} G_F\text{GeV}^2 \,,\nonumber\\
 &&B(\Lambda_c^+ \to \Lambda^{\prime 0} \pi^+)=(16.7\pm 0.9)10^{-2} G_F\text{GeV}^2,
 \end{eqnarray}
with the correlation of $R=0.78$.
 
 According to the Eqs.~\eqref{f1}, \eqref{f2} and \eqref{f3}, the isospin breaking effects caused by the mixing are given by
\begin{eqnarray}\label{finalMIX}
&&\Delta{\cal B}(\Lambda_c^+ \to \Sigma \pi ) \equiv {\cal B}(\Lambda_c^+ \to \Sigma^0 \pi^+)- {\cal B}(\Lambda_c^+ \to \Sigma^+ \pi^0)= (3.8 \pm 0.5)\times 10^{-4}\,,\nonumber\\
&&\Delta \alpha(\Lambda_c^+ \to \Sigma \pi ) \equiv \alpha(\Lambda_c^+ \to \Sigma^0 \pi^+)- \alpha(\Lambda_c^+ \to \Sigma^+ \pi^0)= (-1.6\pm 0.7)\times 10^{-2}\,,
\end{eqnarray}
which are consistent with the current experimental data, given by $(5\pm 12)\times 10^{-4}$ and $(16\pm22)\times 10^{-2}$~\cite{pdg}, respectively. Since the data are also consistent with zero, it is clear that future experiments with higher accuracy are needed.

%\textcolor{red}{
The mixing effects in $\Lambda_c^+ \to \Sigma \pi$ are in the first order of $\theta$\,, while those in $\Lambda_c^+\to \Sigma^0 e^+ \nu_e$ and $\Lambda_b^0 \to \Sigma^0 J/\psi$~\cite{Dery:2020lbc}  in  the second one. Clearly, the experiments in $\Lambda_c^+ \to \Sigma \pi$ are more promising for searching the $\Sigma^0-\Lambda^0$ mixing. 
%}
\section{Conclusions}
We have analyzed the mass differences of the octet baryons.
We have identified the contribution from QED in the baryons mass splittings.
From the baryon masses we have found that the mixing angle between $\Sigma^0$ and $\Lambda^0$ is $\theta = (2.07\pm0.03)\times 10^{-2}$. 
The possibility of observing such a mixing in the $\Lambda_c^+$ decays has been discussed.
 In particular, we note that the decay channel of $\Lambda_c^+ \to \Sigma^0 e^+ \nu_e$ is forbidden if $\theta=0$. With the mixing, the decay branching ratio and up-down asymmetry  in $\Lambda_c^+ \to \Sigma^0 e^+ \nu_e$ are given by
$(1.5\pm 0.2) \times 10^{-5}$ and $-0.86\pm 0.04$, respectively.
In the nonleptonic decays, we have demonstrated that the mixing causes slight differences for the physical observables
between  $\Lambda_c^+\to \Sigma^0\pi^+$ and $\Lambda_c^+\to \Sigma^+\pi^0$. Explicitly, we have shown that $\Delta{\cal B}(\Lambda_c^+ \to \Sigma \pi) =  (3.8 \pm 0.5)\times 10^{-4}$ and $\Delta \alpha(\Lambda_c^+ \to \Sigma \pi)= (-1.6\pm 0.7)\times 10^{-2}$,
respectively. Future experimental searches for ${\cal B}(\Lambda_c^+ \to \Sigma^0 e^+ \nu_e)$, $\Delta {\cal B}(\Lambda_c^+ \to \Sigma \pi)$, 
and $\Delta \alpha(\Lambda_c^+ \to \Sigma \pi)$ are recommended. Non-vanishing values of these observables can be the evidences of the $\Sigma^0-\Lambda^0$ mixing.

\section*{ACKNOWLEDGMENTS}
This work was supported in part by National Center for Theoretical Sciences and
MoST (MoST-107-2119-M-007-013-MY3).


\begin{thebibliography}{99}
	%\cite{Coleman:1961jn}
	\bibitem{Coleman:1961jn} 
	S.~R.~Coleman and S.~L.~Glashow,
	%``Electrodynamic properties of baryons in the unitary symmetry scheme,''
	Phys.\ Rev.\ Lett.\  {\bf 6}, 423 (1961).
	\bibitem{latticeQCD2} 
	R.~Horsley {\it et al.} [QCDSF and UKQCD Collaborations],
	%``Isospin breaking in octet baryon mass splittings,''
	Phys.\ Rev.\ D {\bf 86}, 114511 (2012).
	\bibitem{latticeQCD1} 
	P.~E.~Shanahan, A.~W.~Thomas and R.~D.~Young,
	%``Strong contribution to octet baryon mass splittings,''
	Phys.\ Lett.\ B {\bf 718}, 1148 (2013).
	\bibitem{latticeQCDQED2} 
	S.~Borsanyi {\it et al.} [Budapest-Marseille-Wuppertal Collaboration],
	%``Isospin splittings in the light baryon octet from lattice QCD and QED,''
	Phys.\ Rev.\ Lett.\  {\bf 111}, no. 25, 252001 (2013).
	\bibitem{latticeQCDQED3} 
	S.~Borsanyi {\it et al.},
	%``Ab initio calculation of the neutron-proton mass difference,''
	Science {\bf 347}, 1452 (2015).
	\bibitem{latticeQCDQED1} 
	R.~Horsley {\it et al.},
	%``Isospin splittings of meson and baryon masses from three-flavor lattice QCD + QED,''
	J.\ Phys.\ G {\bf 43}, no. 10, 10LT02 (2016).

	\bibitem{Isgur:1979ed} 
	N.~Isgur,
	%``Isospin violating mass differences and mixing angles: the role of quark masses,''
	Phys.\ Rev.\ D {\bf 21}, 779 (1980)
	Erratum: [Phys.\ Rev.\ D {\bf 23}, 817 (1981)].
	\bibitem{theo1} 
	A.~Blotz, D.~Diakonov, K.~Goeke, N.~W.~Park, V.~Petrov and P.~V.~Pobylitsa,
	%``The SU(3) Nambu-Jona-Lasinio soliton in the collective quantization formulation,''
	Nucl.\ Phys.\ A {\bf 555}, 765 (1993).
	\bibitem{theo5} 
	S.~A.~Coon,
	%``The u - d quark mass difference and nuclear charge symmetry breaking,''
	Nucl.\ Phys.\ A {\bf 689}, 119 (2001).
	\bibitem{theo4} 
	L.~Durand and P.~Ha,
	%``Electromagnetic corrections to baryon masses,''
	Phys.\ Rev.\ D {\bf 71}, 073015 (2005)
	Erratum: [Phys.\ Rev.\ D {\bf 75}, 039903 (2007)].
	\bibitem{theo6CG} 
	P.~Ha,
	%``Estimates of isospin breaking contributions to baryon masses,''
	Phys.\ Rev.\ D {\bf 76}, 073004 (2007).
	\bibitem{theo3} 
	G.~S.~Yang, H.~C.~Kim and M.~V.~Polyakov,
	%``Electromagnetic mass differences of SU(3) baryons within a chiral soliton model,''
	Phys.\ Lett.\ B {\bf 695}, 214 (2011).

	\bibitem{theo2} 
	G.~S.~Yang and H.~C.~Kim,
	%``Mass splittings of SU(3) baryons within a chiral soliton model,''
	Prog.\ Theor.\ Phys.\  {\bf 128}, 397 (2012).
	
		\bibitem{Horsley:2014koa} 
	R.~Horsley {\it et al.},
	%``Lattice determination of Sigma-Lambda mixing,''
	Phys.\ Rev.\ D {\bf 91}, no. 7, 074512 (2015).
	\bibitem{Ablikim:2019zwe} 
M.~Ablikim {\it et al.} [BESIII Collaboration],
%
Phys.\ Rev.\ D {\bf 100}, no. 7, 072004 (2019)
%	arXiv:1905.04707 [hep-ex].
	\bibitem{pdg}
M. Tanabashi {\it et al.} [Particle Data Group],
%``Review of Particle Physics,''
Phys. Rev. D {\bf 98}, 030001 (2018).
	
	\bibitem{georgi}
H.~Georgi, \emph{Lie Algebras in Particle Physics}, Hachette UK, (1999).



	
	
	\bibitem{Mixing_Eq} 
	A.~Gal,
	%``Comment on "Lattice determination of $\Sigma$-$\Lambda$ mixing",''
	Phys.\ Rev.\ D {\bf 92}, no. 1, 018501 (2015)
	\bibitem{Mixing_El} 
	R.~H.~Dalitz and F.~Von Hippel,
	%``Electromagnetic $\Lambda-\sigma^0$ mixing and charge symmetry for the $\Lambda$-hyperon,''
	Phys.\ Lett.\  {\bf 10}, 153 (1964).
	\bibitem{Weinberg:1977hb} 
	S.~Weinberg,
	%``The Problem of Mass,''
	Trans.\ New York Acad.\ Sci.\  {\bf 38}, 185 (1977).
	\bibitem{Franklin:1981rc} 
	J.~Franklin, D.~B.~Lichtenberg, W.~Namgung and D.~Carydas,
	%``Wave Function Mixing of Flavor Degenerate Baryons,''
	Phys.\ Rev.\ D {\bf 24}, 2910 (1981).
	\bibitem{Geng:2019xbo} 
	C.~Q.~Geng, C.~W.~Liu and T.~H.~Tsai,
	%``Asymmetries of anti-triplet charmed baryon decays,''
	Phys.\ Lett.\ B {\bf 794}, 19 (2019).

\bibitem{Dery:2020lbc} 
A.~Dery, M.~Ghosh, Y.~Grossman and S.~Schacht,
%``SU(3)$_F$ Analysis for Beauty Baryon Decays,''
arXiv:2001.05397 [hep-ph].
\end{thebibliography}
\end{document}